# Reconfigurable Optical Networks with Self-Tunable Transceivers: Implementation Options and Control

Michael H. Eiselt

ADVA SE, Maerzenquelle 1-3, 98617 Meiningen, Germany

meiselt@adva.com

**Abstract** *This paper reviews methods for autonomous tuning of optical transceivers, based on an overhead management channel between the modules on both sides of the link. Different implementation options for the tuning principle, as well as for the tunable laser are introduced.*

**Introduction**

The increasing deployment of 5G mobile stations also places a challenge on the optical network supporting the features of 5G. A section of the optical network with a particularly rapid growth is the fronthaul between the central office and the radio unit. Depending on the functional split, data rates of 10 Gbit/s and above need to be transported to each of typically three antenna sectors. With a high antenna density, this part of the network is highly cost-sensitive and requires an implementation supporting simple operation.

These requirements can be supported by autonomously tunable transceivers. ITU-T Recommendation G.698.4[1] (ex: G.metro) has been developed to allow interoperability between transceivers of different vendors. Since publication of this standard, several solutions have been marketed by different vendors, all of them only partly implementing the specifications of the Recommendation.

This paper will first describe the system and network requirements and discuss implementation options of tunable lasers as well as two methods to tune the laser to the target frequency.

**Network architecture**

A typical architecture of the fronthaul network connects the central office (CO) with a number of remote radio units (RU) by point-to-point links. While, in a fiber-rich environment, the connections can run on individual fiber pairs, the increasing number of radio units requires sharing of the transmission fiber by multiple connections, using wavelength division multiplexing (WDM). The fiber layout can form a tree structure with a trunk fiber connecting the CO to a remote node, to which all RUs are connected via drop fibers. As the connections between CO and RU use individual wavelengths, the remote node can include a wavelength demultiplexer to route the wavelength channels. Alternatively, the fiber can be deployed in a drop line structure, passing the radio units and dropping/adding individual wavelengths or sets of wavelengths to/from each radio unit. The drop line structure can be expanded to a "horseshoe" type network by adding a second CO to the end of the line, such that each radio unit can be connected to two central offices, using the same wavelength on separate sections of the ring. This provides redundant connectivity in case of a fiber interruption. In any case, it is important to note that there is no direct connectivity between different radio units, but only between the central office and each radio unit.

**Wavelength agnostic transceivers**

To keep the operational complexity of the front haul system low, tunability of the transceivers at least at the radio units is required. At the same time, however, the cost of these transceivers, which are expected to be deloyed in large numbers, must be kept low. This resulted in the attempt to move control functionality from the individual RU transceivers to a shared function in the central office. One important aspect is here the tuning of the transmitter to a target frequency, used for the RU-to-CO transmission, and to maintain a stable frequency. The target tuning of the transmitter can be performed in two ways. Either, (1) the transceiver is characterized, before shipping to the customer, generating a calibration table with tuning parameters for all operating frequencies or, (2) upon turn-on, the transmitter sweeps over the operating frequency range and expects a feedback from the receiver if the correct frequency has been achieved. After tuning, a stable transmission frequency needs to be maintained. While ITU-T standards do not specify a parameter for frequency stability, the "maximum spectral excursion"[2] limits the modulation bandwidth (including chirp) as well as the central frequency deviation. With a low-chirp modulation, a maximum spectral excursion of $\pm 12.5$ GHz, as specified for many DWDM applications, leaves a central frequency tolerance of approximately $\pm 7.5$ GHz for the tunable laser.

**Management channel**

For communication between the central office and the radio unit, to remotely control the transceiver or to monitor the equipment, a 50-kbit/s overhead channel has been defined in G.698.4. As opposed to in-band signalling channels, which are available for some signal formats (e.g. OTN, CPRI), or data packets, which could be inserted in an Ethernet link, this management channel is independent of the carried data and is already available before the main communication link is established. The overhead channel can be realized by modulating the average power of the (modulated or unmodulated) data channel and can therefore be detected with a slow direct-detection receiver. To avoid interference between the management channel and the data channel, the spectrum of the management channel is required to be limited below ~100 kHz. At the same time, low-frequency components of the management channel, resulting in slow power variations, should be avoided to enable amplification of the optical channel in an EDFA without incurring cross talk to other channels. Therefore, Manchester coding[3] is used for the overhead channel, limiting the spectral content of the 50-kbit/s channel between approximately 10 and 90 kHz.

The sensitivity of the management channel, modulated on to a 2.5-Gbit/s or a 10-Gbit/s data channel can be as low as -40 dBm for a bit-error-rate (BER) of $10^{-6}$, about 10 dB below the sensitivity of the data channel[4]. For error correction, G.698.4 specifies simple error correcting Hamming codes, which yield an average time of 17 hours between lost messages, even with a BER of $5 \cdot 10^{-6}$.

**Initial frequency tuning**

As discussed above, there are two principally different methods to steer the transmitter laser to the correct frequency. If the transceiver is properly calibrated, such that it can tune to a requested frequency, the CO can send the target frequency via the communication channel and the correct tuning parameters are taken from the transceiver calibration table (method (1)). For cost savings, however, the laser calibration procedure might be reduced, such that only few operating points are tested, and the laser cannot be directly tuned to a requested frequency. For this case, G.698.4 provides method (2), where the RU laser sweeps over the transmission band. The laser frequency is filtered by the multiplexer and demultiplexer in the link and reaches the associated CO transceiver only, when the emission frequency matches the mux/demux port frequency. When this happens, the CO sends a message to the RU and the RU stops sweeping. To avoid optical cross talk from the sweeping laser, which might be suppressed only by the multiplexer filter, onto the operating channels, the sweeping laser operates at a reduced power. The power reduction is, in turn, based on an estimation of the link loss between the CO and the tuning RU, such that even high loss differences between the tuning and the operating channels can be accommodated. It should be mentioned that, while this tuning method is provided for in the standard, currently all commercial products use the direct tuning method (1).

**Tunable lasers**

A wide-band tuning of lasers can be achieved based on a variety of methods. The general laser principle is based on a resonance within the laser cavity such that only that particular frequency is lasing, for which the optical round-trip phase in the laser cavity is a multiple of 360 degrees and the gain is sufficient to overcome the facet losses. For short laser cavities, with a length below ~1 µm, only one of the resonance freqencies falls into the bandwidth of the gain medium, which can be between ~10 and ~100 nm. Changing the length of the cavity would then result in a tuning of the laser frequency. Vertical cavity surface emitting lasers (VCSELs) with one movable facet based on a micro electro-mechanical system (MEMS) use this tuning principle[5],[6]. The advantage of these lasers is that the laser is intrinsically single-mode and only a single parameter is required for frequency tuning. For these types of laser, the sweeping tuning method (2) is well suited, and a calibration of the laser tuning is not necessary.

If the laser cavity is longer, as for waveguide lasers, the spacing between resonance frequencies is closer and a filter must be provided in the laser cavity to avoid multi-mode lasing. One implementation of the filter is a waveguide Bragg grating on one side of the laser gain section, while on the other side all frequencies are reflected, e.g. at the waveguide facet. The Bragg grating is typically temperature-tuned and reflects a bandwidth, which should be narrower than the resonance frequency (mode) spacing of the cavity. For a fine tuning of the cavity resonance frequencies, an additional phase section is required, such that for frequency tuning two parameters need to be controlled. For most materials, the typical frequency shift of a Bragg grating, based on the temperature coefficient of the refractive index, is on the order of 0.1 nm/K. For a tuning over a bandwidth of 20 nm, a temperature shift of ~200 K is required, which is only feasible in a few material systems. A structure with a gain element in InP, coupled to a waveguide with Bragg grating in a Polymer substrate, has shown to be tunable over more than 25 nm[7]. A slightly higher temperature coefficient of the Polymer material together with a low thermal conductivity enabled a tuning efficiency of 0.52 nm/mW.

Less thermal variation is required in a structure, where Bragg gratings with periodic reflection spectra are provided on both sides[8] or, after a Y-shape coupler[9], on one side of the gain section. With a slightly different free spectral range of the two gratings a Vernier method is used to tune the laser, still requiring a phase section for fine tuning of the cavity resonance. This increases the frequency tuning parameters to three, in addition to the injection current into the gain section. As this tuning principle allows a tuning over several tens of nanometers with a limited tuning power, this type of tunable lasers can often be found in commercial applications. Due to the higher tuning complexity, however, in-factory calibration of the laser tuning is required and the sweeping method (2) of the transmitter tuning loses its advantage.

**Further work**

Current implementations of self-tunable transceivers are mostly not based on the specifications in G.698.4, but follow proprietary concepts. Only some commercial transceiver products follow the wavelength plan of the Recommendation. Of course, this leads to a lack of interoperability between different solutions.

G.698.4 is built on an asymmetric link architecture: multiple single channel interfaces in the RUs are connected to a multi-channel interface in the CO, such that shared wavelength control becomes feasible. Since the development of G.698.4, however, an additional requirement has as arisen: operators prefer a pair-wise self-tuning of the interfaces. Implementations already exist, where one of the transceivers acts as the primary node and gets input from the network management system, on which channel to operate. It then conveys this information, via the management channel, to the dependent node, which then also performs the appropriate tuning. Ideally, however, after turning up the transceivers on both sides of the link, both modules tune to the wavelengths that enable communication through the multiplexer / demultiplexer to the transceiver on the other side of the link, without intervention from the operator. This self-tuning will simplify operation of the transmission system and will probably find applications also in other parts of the optical network.

**Summary**

Self-tuning transceivers have been introduced in front-haul networks to reduce equipment cost and to simplify operations. Since the introduction of ITU-T Recommendation G.698.4, commercial implementations have only partly followed the specification. Now, modifications of the self-tuning principle are under discussion to allow a pairwise self-tuning between transceivers, which might then find applications also in other parts of the network.